\documentclass[epj]{svjour}
\usepackage{graphics}
\begin{document}
\title{The $Dip$ Effect under Integer Quantized Hall Conditions}
\author{S. Erden Gulebaglan\inst{1}, S. B. Kalkan\inst{2}, S. Sirt\inst{2}, E. M. Kendirlik\inst{2} \and A. Siddiki\inst{2}
}                     
%
%
\institute{Y\"{u}z\"{u}nc\"{u} Y{\i}l University, Physics Department, Faculty of Sciences, 65080 Van, Turkey \and Istanbul University, Physics Department, Faculty of Sciences, 34134-Vezneciler-Istanbul, Turkey}
\date{Received: date / Revised version: date}
%
\abstract{
In this work we investigate an unusual transport phenomenon observed in two-dimensional electron gas under integer quantum Hall effect conditions. Our calculations are based on the screening theory, using a semi-analytical model. The transport anomalies are \emph{dip} and overshoot effects, where the Hall resistance decreases (or increases) unexpectedly at the quantized resistance plateaus intervals. We report on our numerical findings of the \emph{dip} effect in the Hall resistance, considering GaAs/AlGaAs heterostructures in which we investigated the effect under different experimental conditions. We show that, similar to overshoot, the amplitude of the dip effect is strongly influenced by the edge reconstruction due to electrostatics. It is observed that the steep potential variation close to the physical boundaries of the sample results in narrower incompressible strips, hence, the experimental observation of the dip effect is limited by the properties of these current carrying strips. By performing standard Hall resistance measurements on gate defined narrow samples, we demonstrate that the predictions of the screening theory is in well agreement with our experimental findings.
} 
\authorrunning{S. Erden Gulebaglan}
\titlerunning{The dip effect}
\maketitle
\section{Introduction}
\label{intro}
After the path paving discovery of integer quantum Hall effect (IQHE)~\cite{Klitzing} in two-dimensional magneto-transport, many experiments have been performed to unreveal the transport properties of the low dimensional charge systems in the presence of strong perpendicular magnetic fields. In the presence of an external $B$ field the density of states of the system is quantized, namely the Landau quantization, and the energy of the Landau levels are given by $E_{\rm n}=\hbar \omega_{\rm c}(n+1/2)$. Here $\omega_{\rm c} = eB/m$ is the cyclotron frequency of an electron with an effective mass $m^*$ ($m^* \approx0.067m_{\rm e}$, $m_{\rm e}$ is the electron mass) and $n$ is the Landau index. IQHE observed in two-dimensional electron gas (2DEG) subject to high magnetic fields, is well understood in terms of screening effects, where the system is composed of compressible and incompressible strips~\cite{Siddiki04:195335}. In the compressible regions, the Fermi energy is pinned to one of the Landau levels with high density of states. In contrast, the Fermi energy is in between quantized Landau levels in the incompressible regions, where no density of states is available. The incompressible regions behave like an insulator, whereas the compressible regions can be considered as a metal~\cite{Chklovskii92:4026,Lier,Siddiki:2003,Guven}. The induced Hall potential drops across the incompressible strips, by the virtue of their scattering free transport properties. This interaction based explanation assumes a local relation between the electric fields and current densities $\vec{j}(\vec{r})$, utilizing the Ohm's law, if the local conductivity tensor elements are provided. The conductivities are obtained from the self-consistent Born approximation~\cite{Ando}. We introduce the filling factor $\nu=2\pi l^2n_{\rm el}$, together with its local counterpart $\nu(x)=2\pi l^2n_{\rm el}(x)$, where $l=(\hbar/eB)^{1/2}$ is the magnetic length and $n_{\rm el}$, $n_{\rm el}(x)$ are the average and the local electron number densities. The Hall conductivity is quantized $\sigma_H=\frac{h}{e^2}\nu$ and the longitudinal conductivity vanishes at zero temperature, $\sigma_l\rightarrow0$ ~\cite{Siddiki04:195335,Ando}. This leads immediately to the IQHE for the global resistance $R_H\rightarrow(\frac{1}{\nu}\frac{e^2}{h})$, without any localization assumptions for an inhomogeneous 2DEG with filling factors near $\nu=n$ and $n$ being an (even) integer~\cite{Gerhardts}.

Under the typical IQHE conditions, i.e. wide samples ($W=2d>100$ $\mu$m) with low mobility, the Hall resistivity of a 2DEG presents a stepwise behavior. However, an unexpected increase (overshoot) or decrease (dip) at the low or high $B$ side of the Hall resistance plateaus are also reported in many different materials~\cite{Shlimak:2006,Galistu,Shlimak:prb2006,Richter,Griffin,Ramvall,Shlimak:2005}. The increase of the Hall resistance is known as the overshoot, and is usually attributed to impurity effects, similar to IQHE. The main approach~\cite{Shlimak:2006,Shlimak:prb2006} that has been developed to explain the \emph{dip} effect phenomenon also relies on the impurity scattering. Shlimak et al. found that, this effect can be considered to be a manifestation of the oscillating enhancement of the valley splitting due to the exchange interaction~\cite{Shlimak:prb2006}, meanwhile in our previous work we found that the overshoot can be well explained within the self-consistent screening theory~\cite{Siddiki:2010}. In the later work it is reported that, the overshoot depends strongly on temperature, in contrast, it is immune to short-range impurity scattering and the edge effects are irruptive to the resistance overshoot. The findings of the screening theory is in well agreement with the experiments performed on SiGe based structures, where both the temperature and size effects are investigated~\cite{sailer:2010}. In a very recent experimental investigation, the predictions of the screening theory based overshoot effect is successfully tested. It is experimentally shown that, the edge re-construction due to direct Coulomb interactions modify the amplitude and temperature dependency of the overshoot effect~\cite{metin:2013}. In this work we tackle the dip effect utilizing the screening theory and investigate theoretically the sample and temperature dependent properties of this transport anomaly. Then, we also perform experiments on gate defined narrow Hall bars which are induced on an intermediate mobility wafer~\cite{jose:2008,siddiki:2009,siddiki:2010njp}.

This paper is organized as follows. First, we discuss the basics of the screening
theory of the IQHE. In the next step, we calculate Hall resistances numerically and investigate
the dependencies of the \emph{dip} on sample size, impurity strength, disorder, electron temperature and depletion length. Finally, we propose that the \emph{dip} effect can be elucidated by the variation of the edge profile. The last Section is devoted to experimental findings where we show that our model can be well justified at least as a proof of concept.
\section{Self-consistent Calculation Scheme}
\label{sec:1}
The numerical calculation scheme discussed below provides a consistent explanation to the IQHE. A detailed description of the transport calculations can be found in our previous work~\cite{Siddiki04:195335}. We assume that electrons and donors are on the same $x-y$ plane ($z = 0$) and donors are distributed
homogeneously in the interval  $-d < x < d$, where $d$ is the half-width of the sample and we consider only symmetric density profiles and take $b$, therefore the depletion
length $|d-b|$, as a free parameter and the translation invariance
in the $y$ direction. The electrons move in an effective potential that is described by
\begin{equation}
V(x)=V_{\rm bg}(x)+V_{\rm H}(x).
\end{equation}

Here $V_{\rm bg}(x)=-E_0\sqrt{1-(x/d)^2}$ is the background potential and $E_0=2\pi e^2n_0d/\kappa$ is the minimum of the confinement. Furthermore, $en_{\rm 0}$ is the  density of positive background charges (homogeneous) in the Hall system. The Hartree potential
\begin{equation}
V_{\rm H}(x)=\frac{2e^2}{\kappa}\int_{-d}^d dx'K(x,x')n_{\rm el}(x')
\end{equation}
describes the direct Coulomb interaction between the electrons, where the kernel $K(x,x^{'})$ is given by
\begin{equation}
K(x,x^{'})=\ln\left|\frac{\sqrt{(d^2-x^2)(d^2-{x'^{2}})}+d^2-x^{'}x}{(x-x^{'})d}\right|.
\end{equation}

The Kernel solves the Poisson's equation under the given boundary conditions $V(-d)=V(d)=0$, and $e$ is the charge of an electron, together with $\overline{\kappa}$ being an average background dielectric constant.

The electron density is calculated within the Thomas-Fermi approximation (TFA), where exchange and correlation effects and Zeeman splitting are neglected due to usual assumption of spinless particles. Then the electron density reads,
\begin{equation}
n_{\rm el}(x)=\int dED(E)f(E+V(x)-\mu^*),
\end{equation}
where $f(E)$ is the Fermi distribution function and $\mu^*$
is the equilibrium electrochemical potential. Using the self-consistent Born approximation~\cite{Ando}, the density of states $D(E)$ can be calculated in a straightforward manner. Local conductivities and the density of states (DOS) are obtained supposing an impurity potential having a Gaussian form~\cite{Ando}
\begin{equation}
V(r)=\frac{V_{\rm I}}{\pi R^2}\exp(-\frac{r^2}{R^2}),
\end{equation}
where $V_I$ is the impurity strength and $R$ is the distance between the 2DES and the doping layer, known as the spacer thickness. Due to impurity scattering, the Landau levels are broadened in strong magnetic fields. Additionally, the Landau level width is described by
\begin{equation}
\Gamma^2=4\pi n_{\rm I}^2 V_{\rm I}^2/(2\pi l^2)=(2/\pi)\hbar\omega_c\hbar/\tau,
\end{equation}
where $\tau$ is the momentum relaxation time and $n_{\rm I}$ is the number density of the impurities. For further references it is convenient to characterize the impurity strength by the dimensionless
ratio $\gamma =\Gamma /\hbar \omega_{\rm c}$ ($B$=10 T for GaAs) and the strength parameter reads~\cite{Siddiki04:195335}
\begin{equation}
\gamma_{\rm I}=[(2n_{\rm I}V_0^2m^*/\pi \hbar^2)(1.73 ~{\rm meV})]^{1/2}.
\end{equation}

\subsection{The local conductivity model}
\label{sec:2}
We utilize the linear local relation $\vec{j}(\vec{r})=\hat{\sigma}(\vec{r})\vec{E}(\vec{r})$ (i.e. the local Ohms law), where $\hat{\sigma}(\vec{r})$ is a position-dependent conductivity tensor and $\vec{E(\vec{r})}$ is the electric field. For the sake of consistency, we assumed a translational invariance in $y-$direction and employed the equation of continuity to obtain the local fields and current densities as $E_y=(x)=E_y^0$ and $j_x=0$, respectively. The remaining components are then $j_y(x)=\frac{1}{\rho_\ell(x)}E_y^0$ and $E_x(x)=\frac{\rho_H(x)}{\rho_\ell(x)}E_y^0$. Once the local conductivity tensor entries are obtained from the local electron density within the SCBA, then the local electric fields
can be obtained from the self-consistent potential. The above equations describe local transport properties in a self-standing
manner.

For a given fixed applied current $I =\int_{-d}^d dx j_y(x)$ the electric
field component along the Hall bar is calculated as
\begin{equation}
E_y^0=I[\int_{-d}^d dx\frac{1}{\rho_\ell}]^{-1},
\end{equation}
hence, the Hall voltage is
\begin{equation}
V_{Hall}=\int_{-d}^d dx E_x(x).
\end{equation}

The global measurable quantities such as the Hall and the longitudinal resistances are then obtained as $R_H=\frac{V_{Hall}}{I}$
and $R_\ell=\frac{2dE_y^0}{I}$. It is important to note that, the components of conductivity tensor at integer filling factors read $\sigma_\ell=\rho_\ell=0$ and $\sigma_H=1/\rho_H=\frac{e^2}{h}\nu$ as discussed by Gerhardts and his co-workers~\cite{Siddiki04:195335,Guven}.

\section{Results and discussion}
In this section, we employ the above explained calculation scheme to investigate the dip effect, which is observed at the low or high $B$ field end of a quantized Hall plateau. First, we discuss the effect of sample size where we show that the dip effect tends to disappear at narrower samples, in contrast to overshoot effect~\cite{metin:2013}. Next the effect of disorder is investigated from short-range impurity scattering point of view, namely the influence of level broadening on the dip effect, together with the long-range potential fluctuations aspect. In the later, we impose an external potential with inhomogeneous modulation which is included to our screening calculations. It is observed that the long-range effects are not predominant in determining the amplitude of the dip effect. In the last two investigations we look closely to the temperature effects and the effect of depletion length.

\begin{figure}
\resizebox{0.5\textwidth}{!}{%
  \includegraphics{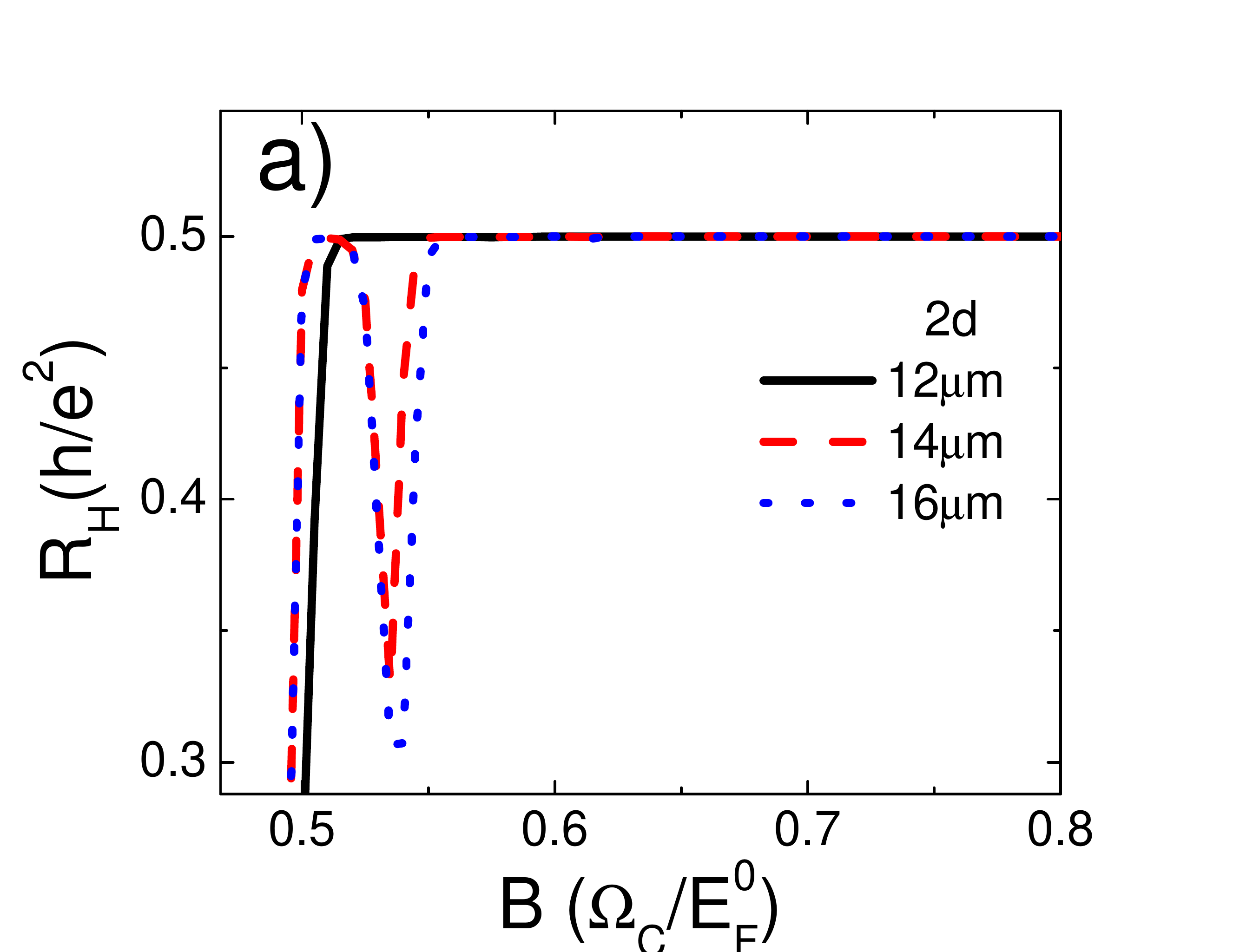}
  \includegraphics{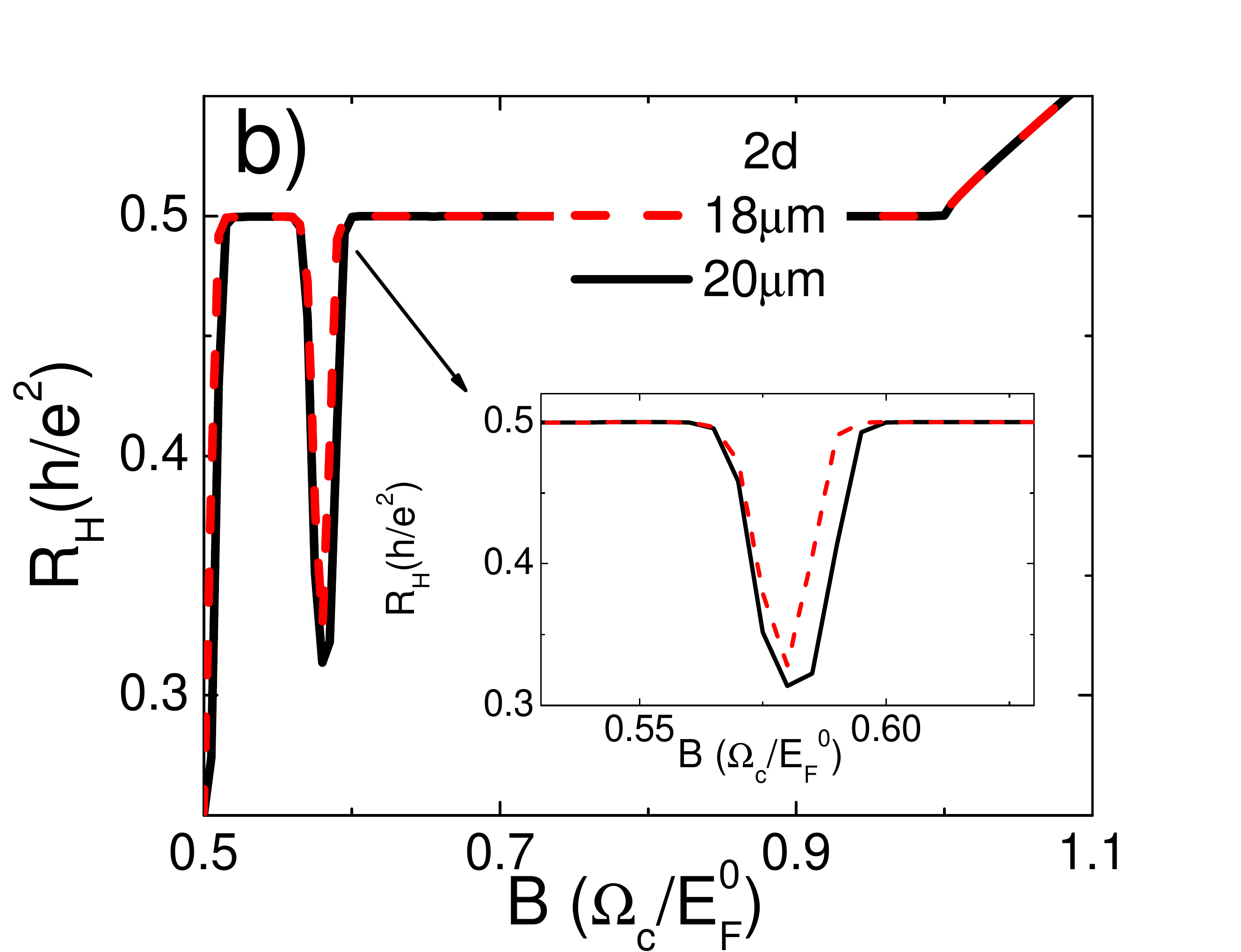}
}
\caption{Hall resistance versus magnetic field, calculated for
different sample widths, a) $2d$ = 12, 14, 16$ \mu$m and b) $2d$ = 18, 20$ \mu$m. The field strength is given in units of $\Omega_c/E_0^F=\hbar\omega_c/E_0^F$, where $E_0^F$ is the Fermi energy at the center of the sample. We fix the impurity scattering parameter $\gamma=0.01$, to eliminate contribution from level broadening. One sees that, dip effect is absent for the 12 $\mu$m sample}
\label{fig:1}       
\end{figure}

We investigate the effect of sample width on the \emph{dip} effect by depicting the filling factor two plateaus calculated within the screening theory of the quantized Hall effect in Fig.~\ref{fig:1}, where temperature is fixed and the plateau widths are compared considering different sample sizes. The impurity parameters and depletion lengths are kept constant. Calculations are done
at $k_BT/E_F^0 = 0.01$ and the donor density is set to be $4\times10^{15}\, \rm{m^{-2}}$ for all samples. We observe that the wider samples present extended plateaus, while the magnetic field is normalized with Fermi energy calculated at the bulk of an homogeneous sample, namely $E_F^0$. The first observation is that, as the sample size increases the amplitude of the \emph{dip} increases. In the addition, the \emph{dip} effect shifts from low-field end of the plateau to intermediate-fields. If the sample is narrow the variation of the confinement potential is stronger, hence the widths of the incompressible strips decrease. In other words, the strength of the confinement potential together with the Lorentz force overcomes the direct Coulomb interaction and as a result the incompressible strip becomes unstable and collapses. However, once the magnetic field strength decreases further the Lorentz force also decreases and the incompressible strip reconstructs hence the Hall resistance recovers its quantized value.

\begin{figure}
\resizebox{0.5\textwidth}{!}{%
  \includegraphics{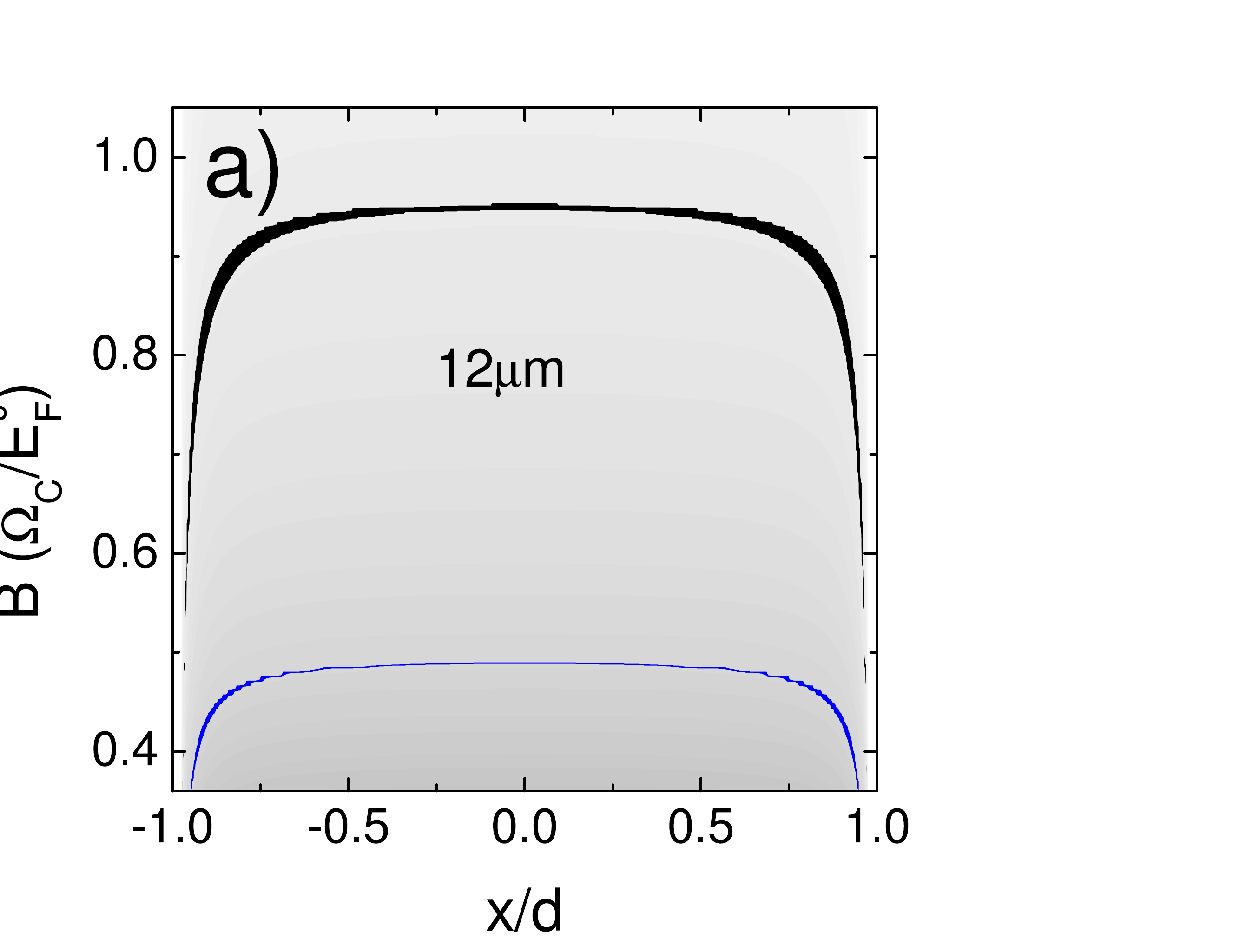}
  \includegraphics{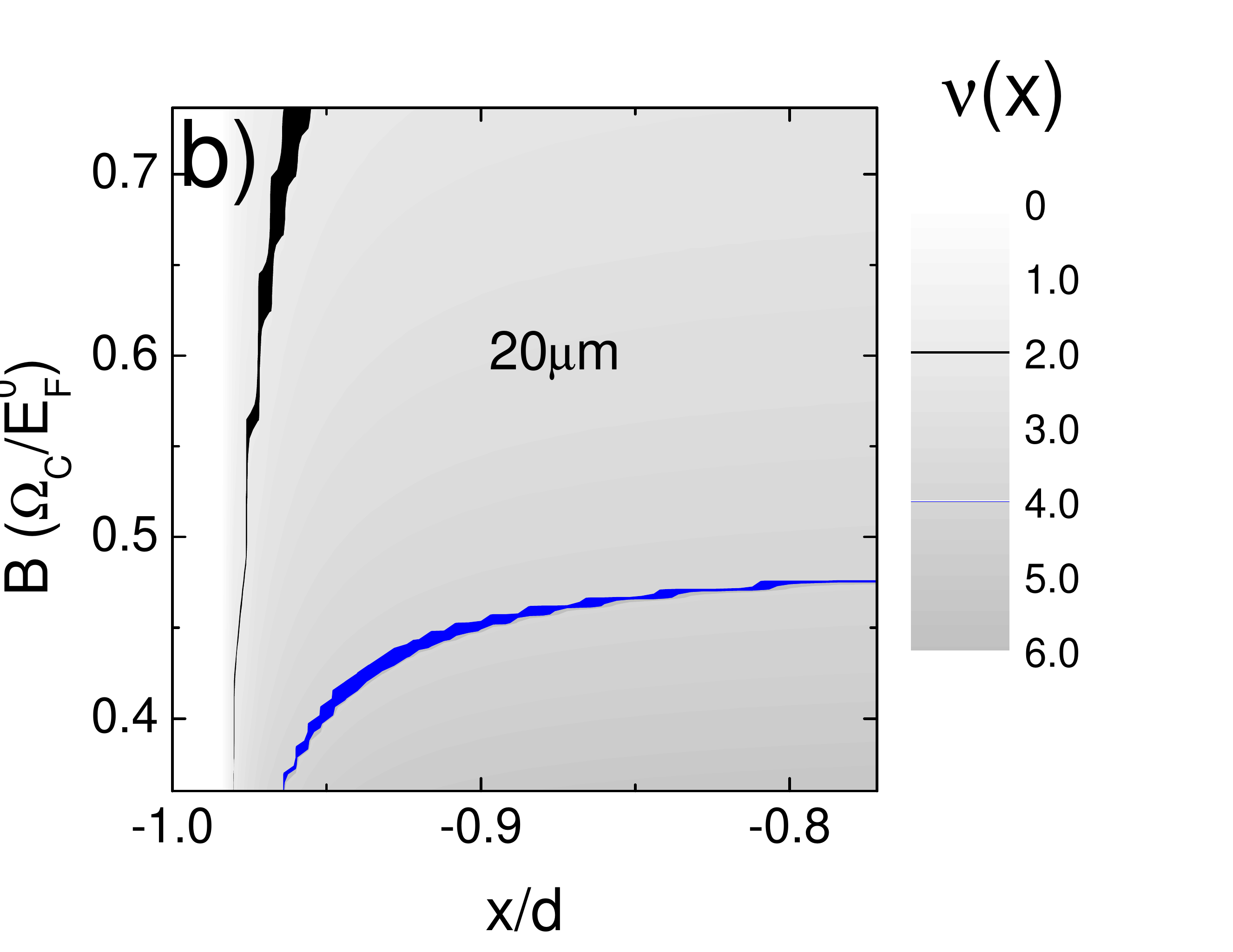}
}
\caption{The local filling factor as a function of normalized lateral coordinate $x/$ and magnetic field $B$. The compressible regions are depicted by gray, whereas the incompressible strips with $\nu(x)=2$ is highlighted by dark (black) and $\nu=4$ lighter (dark blue) colors considering (a) $2d=20$ $\mu$m and (b) $2d=12$ $\mu$m wide samples. Here temperature is set to be $kT/E_F^0=0.01$.}
\label{fig:2}       
\end{figure}
Fig.~\ref{fig:2}a and Fig.~\ref{fig:2}b, plots the local filling factor distribution
as a function of normalized spatial coordinate $x/d$. The black and dark blue regions depict the distribution of the incompressible strips. We will focus on the intervals where the incompressible strips become sufficiently narrow to investigate the resistance \emph{dip}, similar to the overshoot effect. Remember that, the \emph{dip} effect was absent in the 12 $\mu$m sample, but it was present in the 20 $\mu$m wide sample. For the wide sample,
the filling factor $\nu=2$ incompressible strip disappears in the magnetic field range $0.55<B<0.60$, and is reconstructed again outside the corresponding interval, as seen in Fig.~\ref{fig:2}a. Fig.~\ref{fig:2}b shows that, the incompressible strip ($\nu=2$) continue it is existence in 12 $\mu$m sample for all the plateau interval, hence no dip effect is observed. The collapse and reconstruction of the incompressible strip is the main mechanism to observe the \emph{dip} effect.

\begin{figure}
\resizebox{0.5\textwidth}{!}{%
  \includegraphics{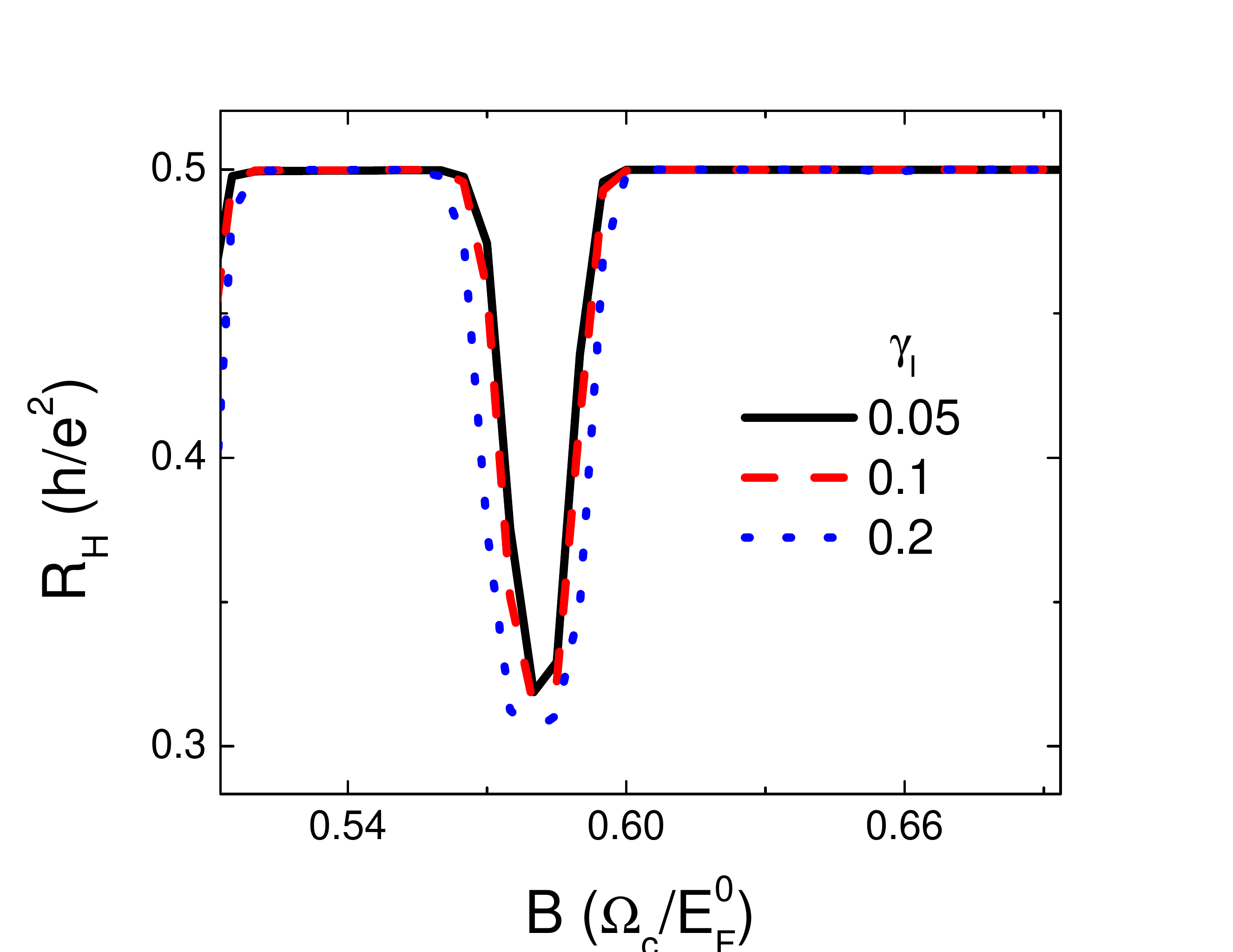}
}
\caption{Hall resistance as a function of magnetic field plotted for different level broadening parameters, $\gamma_I$, considering a 20 $\mu$m wide Hall bar at default temperature $kT/E_F^0=0.01$. The lowest mobility
($\gamma_{\rm I}$=0.2) shows the largest dip effect.}
\label{fig:4}       
\end{figure}

Next we investigate the effect of two parameters, $V_{\rm imp}$ and $N_{\rm I}$, related with the scattering broadening. Note that, these two parameters both effect the level width simultaneously, thereby the
widths of the incompressible strips. Hence, one cannot to distinguish their influence
on the quantized Hall plateaus separately. Characteristic Hall resistances are shown in Fig.~\ref{fig:4}, which are calculated at
default temperature considering different impurity parameters. Namely, the amplitude of the resistance \emph{dip} effect shows a weak dependence on $\gamma_{\rm I}$. We observe that the dip effect becomes largest when the mobility is low,
which also points that our single particle based level broadening calculations are
in the correct direction. We observed that dip effect becomes a little wider when the mobility is low. Such a behavior is easy to understand, since both the energetic and spatial gap between consequent levels shrink, hence the incompressible strips collapse easily. As a result the effect of impurity on the resistance dip is negligible.

In one of our recent work~\cite{Sinem:2012}, we showed that the total potential fluctuates over a spread scale of nearly two-hundred nanometers for the high impurity concentration, however the length scale is found to be approximately few micrometers at the low impurity concentration. To unreveal the effect of long-range fluctuations, we also include an external modulation potential to our screening calculations in addition to the confinement potential and investigate the long range disorder potential as well as its effects on the self-consistent potential. Modulation
potential which we used is defined as; $V_{\rm mod}(x)=V_0\cos{(\frac{ 2 \pi x m_{p}}{2d}})$. Here $m_p$ is the modulation period which is preserves the boundary
conditions. In this study, we vary the amplitude of the modulation potential and consider two different modulation periods regardless of the sample width.

\begin{figure}
\resizebox{0.5\textwidth}{!}{%
  \includegraphics{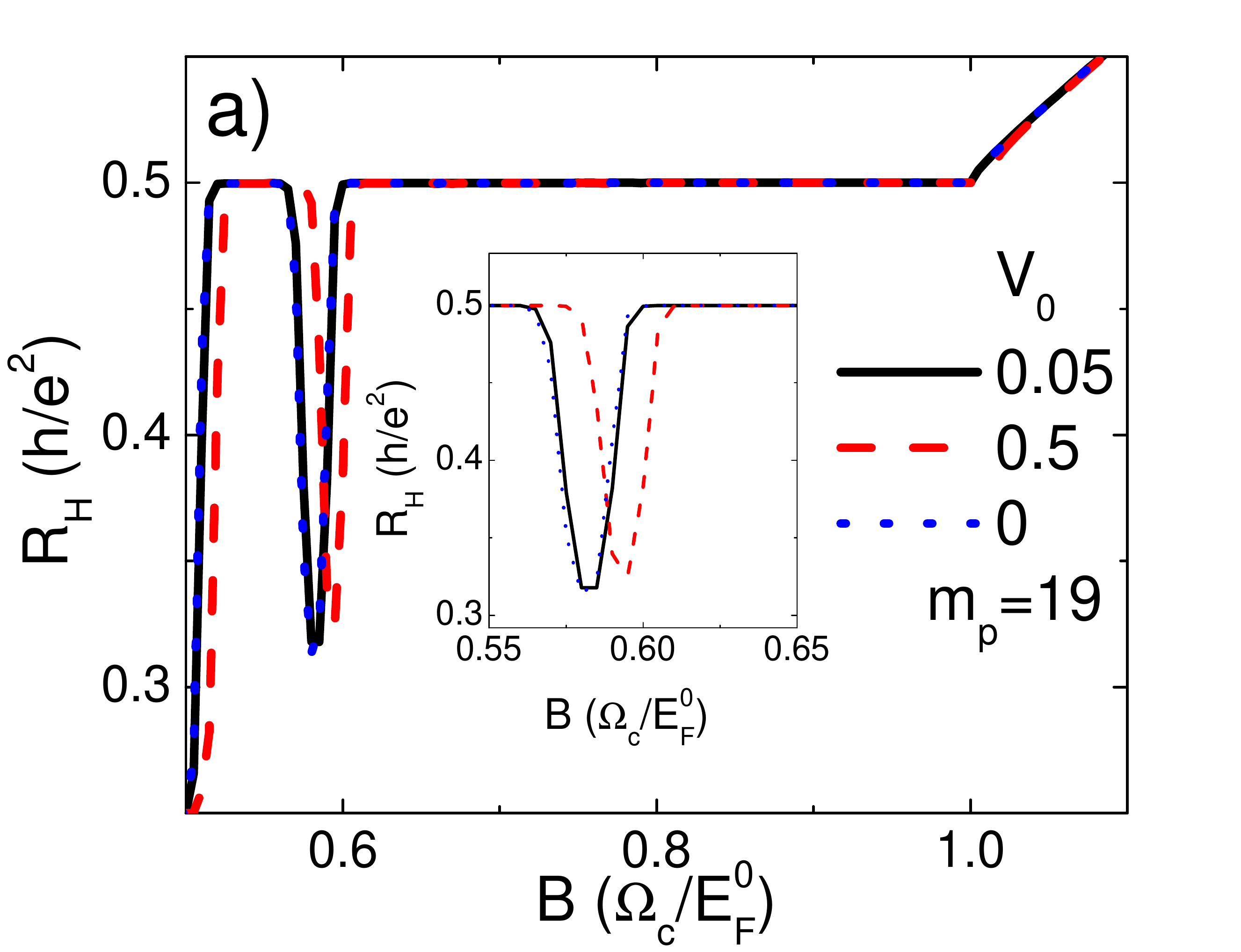}
  \includegraphics{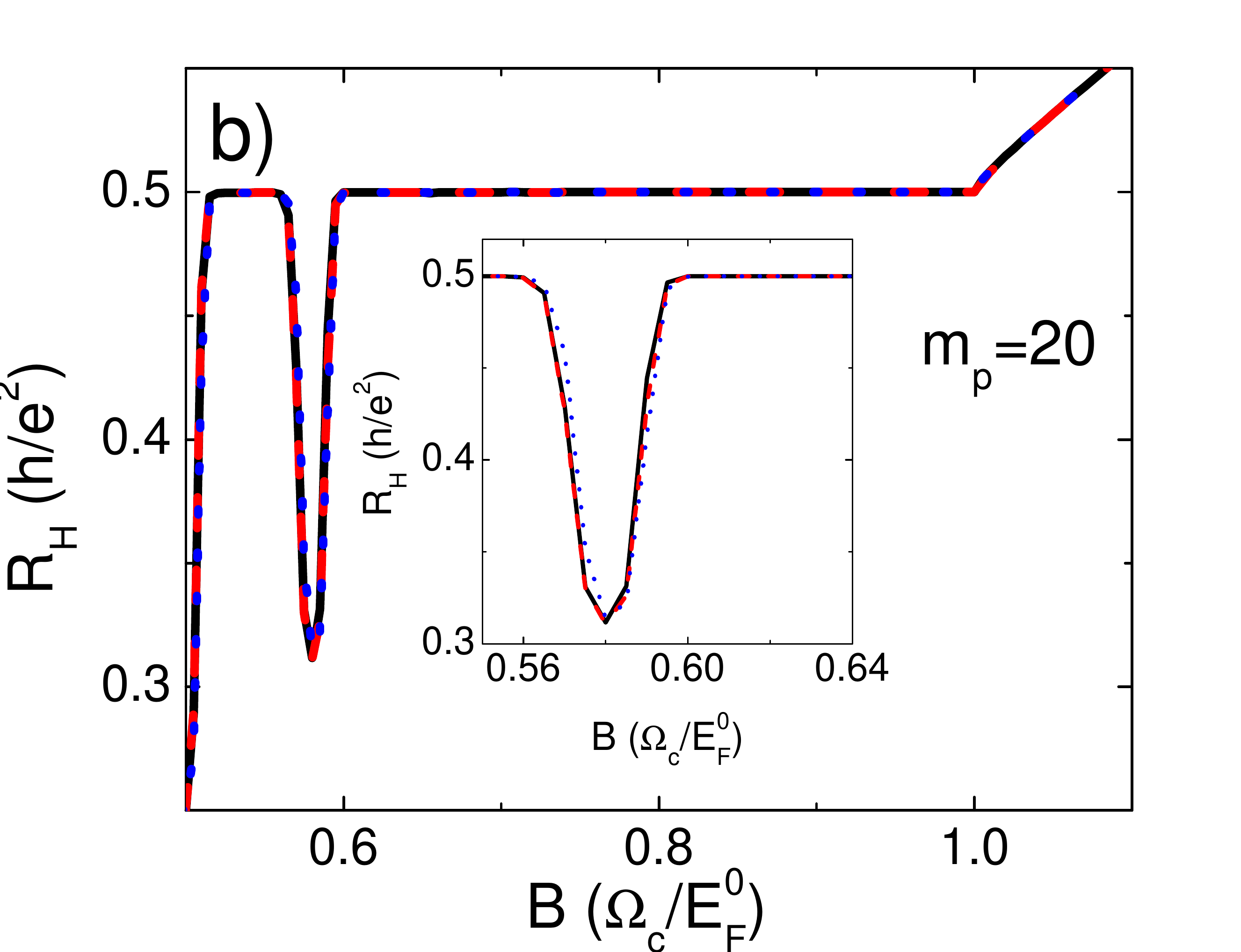}
}
\caption{The calculated the Hall resistance versus magnetic field considering  sample width ($2d = 20 ~\mu$m). The impurity strength and depletion length parameters are kept fixed, whereas the parity of modulation period is odd a) and even b).}
\label{fig:5}       
\end{figure}

Fig.~\ref{fig:5} summarizes our results considering both low and high mobility regimes for two different
$V_{\rm 0}$. When $V_{\rm 0}$ is increased, mobility is reduced. Fig.~\ref{fig:5} depicts the sample size dependency of $\nu = 2$ \emph{dip} effect. We present the self-consistently calculated the Hall resistances, considering different modulation amplitudes $V_{\rm 0}$ for a fixed sample width ($2d$ = 20 $\mu$m) and $m_{\rm p}$ = 19 and  $m_{\rm p}$ = 20. We show that, the dip effect occurs at the high B field side, when $V_{\rm 0}$
is increased, i.e mobility is reduced. If the effect of the thickness is taken into account, the amplitude of the disorder potential is reduced to $\%$ 50 of the Fermi energy. We estimate that the high mobility will be identified by a modulation amplitude of $V_0{\rm 0}/E_F^0 = 0.05$, whereas low mobility corresponds to $V_{\rm 0}/E_F^0 = 0.5$. The odd modulation period of $m_{\rm p}=19$ shifts the \emph{dip} effect toward the high field (see Fig.~\ref{fig:5}a) while the even number of oscillations $m_{\rm p} =20$, does not play an important role when considering large samples (Fig.~\ref{fig:5}b) at least for the case under investigation. We also performed calculations where the modulation period is relatively small ($m_{\rm p}$=5 or 6), however, we observed that such small periods have no significant effect on the dip effect, when considering large samples ($2d$=20 $\mu$m). We also observed that, if the maxima of the modulation potential is at the edge of the sample, the incompressible strip suddenly disappears at a higher magnetic field value, whereas, the edge incompressible strips become larger at the lower field side.

\begin{figure}
\resizebox{0.5\textwidth}{!}{%
  \includegraphics{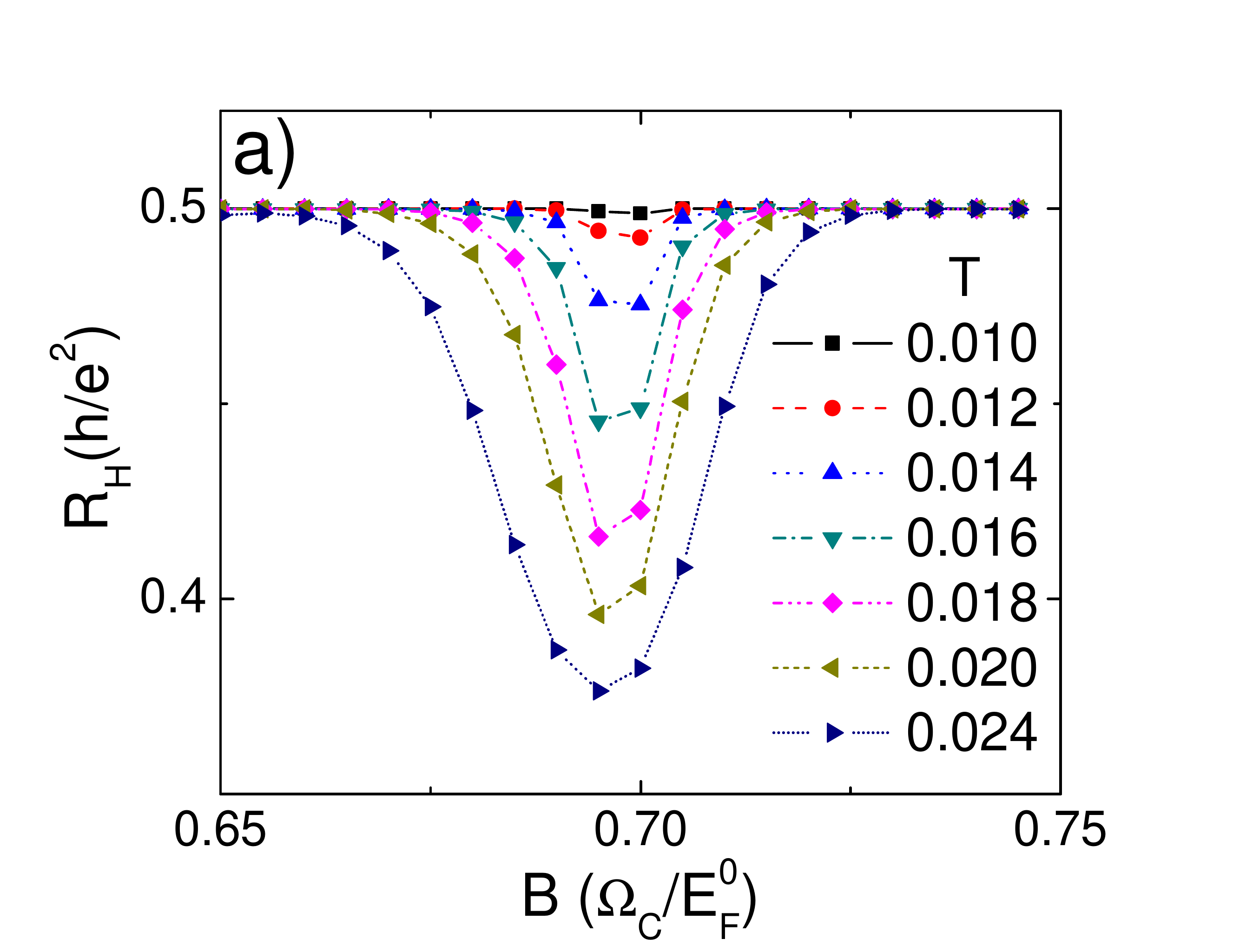}
  \includegraphics{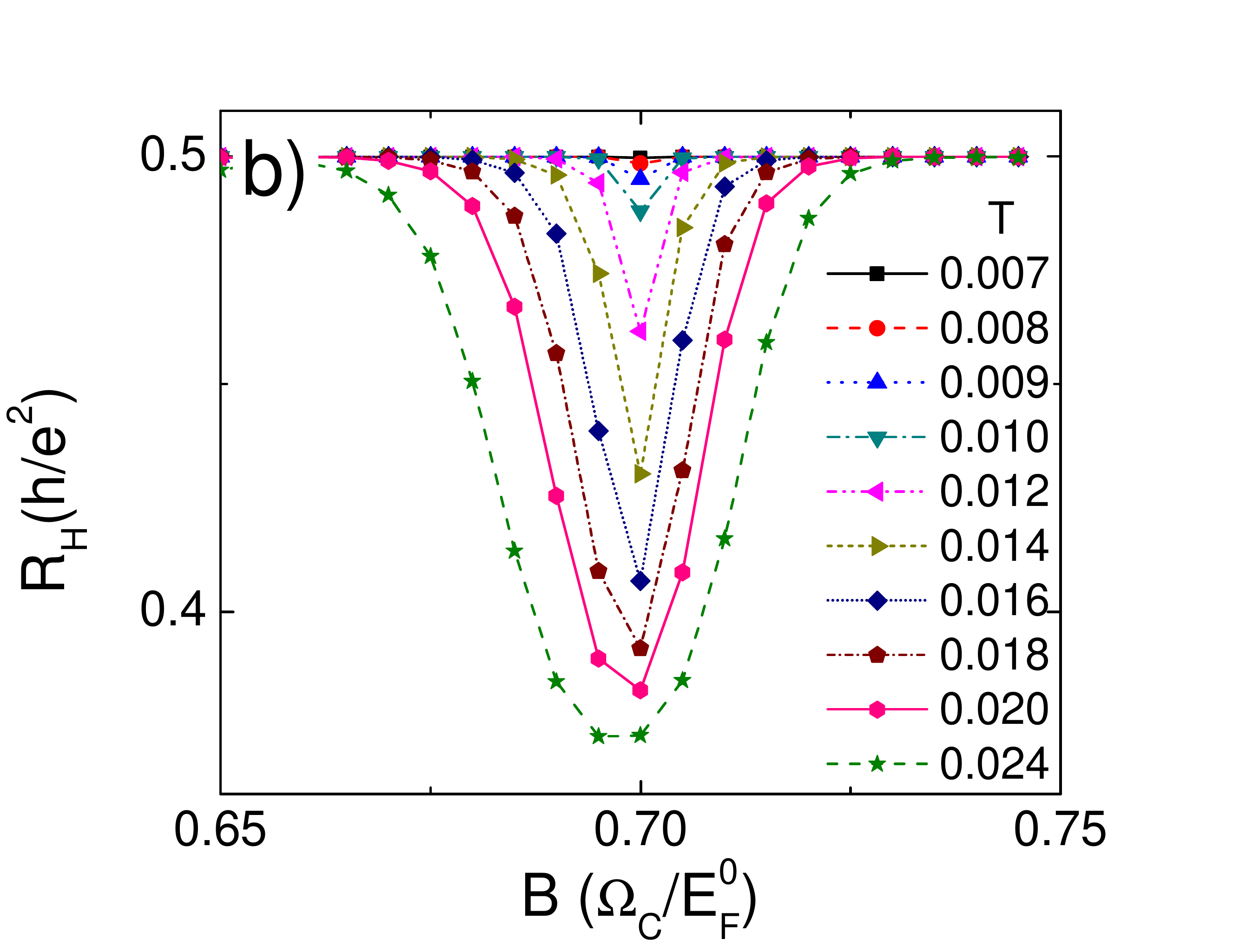}
}
\caption{The calculated Hall resistance as a function of magnetic field investigated at different temperatures considering  18$\mu$m and 20$\mu$m wide samples.}
\label{fig:6}       
\end{figure}

In Figures~\ref{fig:6}a and~\ref{fig:6}b the temperature dependencies are shown considering two different
sample widths. One can clearly see the \emph{dip} effect at
the anticipated magnetic field intervals, where the incompressible strip collapses and is reconstructed. The \emph{dip} effect becomes stronger when increasing the temperature, since the incompressible strip assuming $\nu(x)=2$ collapses easier at higher temperatures. This finding is also consistent with the recent experiments concerning the overshoot effect~\cite{metin:2013}.

\begin{figure}
\resizebox{0.5\textwidth}{!}{%
  \includegraphics{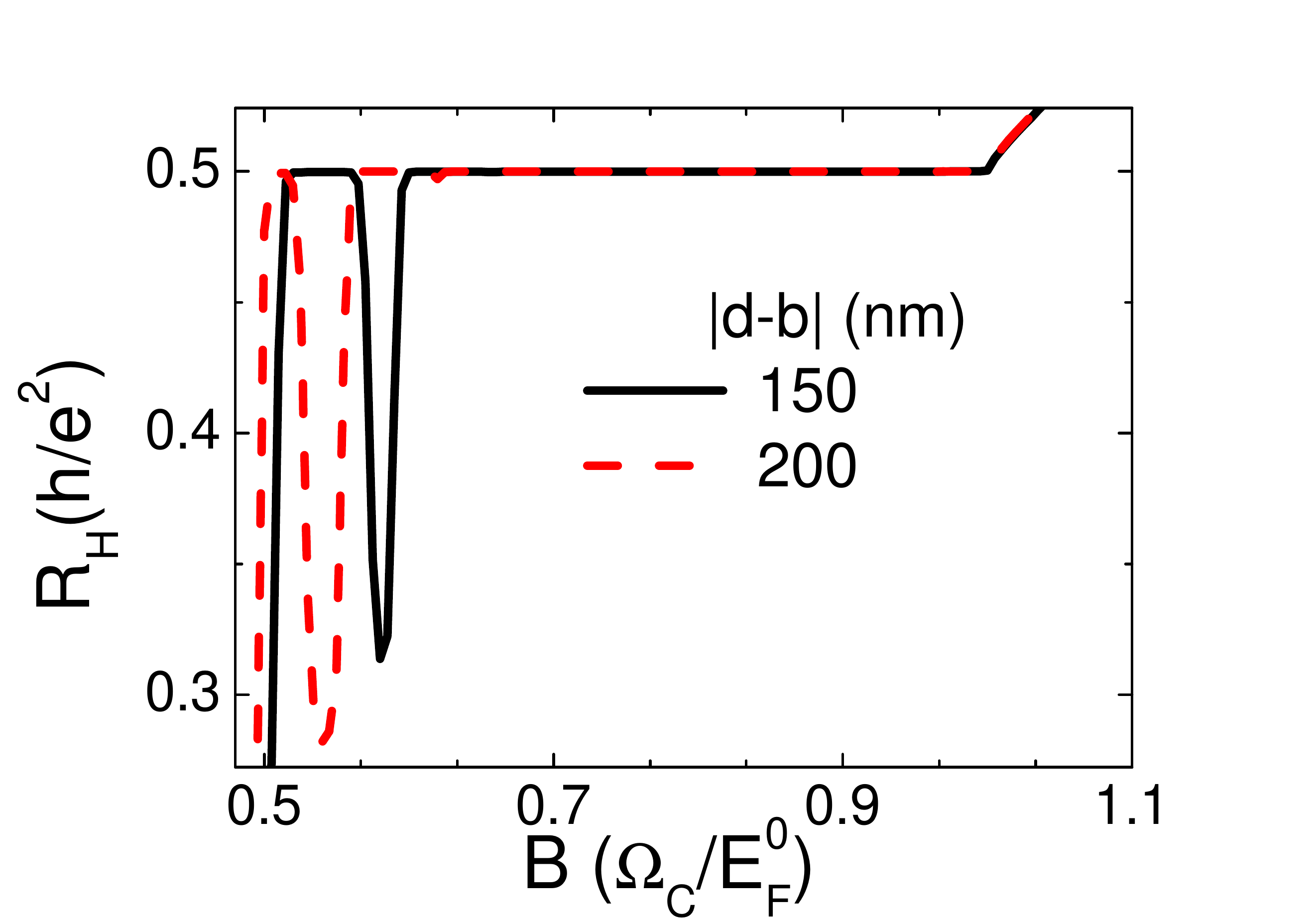}
}
\caption{Hall resistance calculated for different depletion lengths (150 nm and 200 nm) considering a 20$\mu$m wide Hall bar. The temperature is set to be $kT/E_F^0=0.01$.}
\label{fig:7}       
\end{figure}

Finally, we depict the effect of the depletion length on the resistance \emph{dip} as shown in Fig.~\ref{fig:7}. Another important parameter in defining the incompressible strip is the depletion
length. The slope of the confinement potential close to the edges essentially
determines the widths of the incompressible strips ~\cite{Chklovskii92:4026}. We show the $\nu=2$ plateau calculated for two different depletion lengths and see that for the larger depletion the dip effect is more extended, Fig.~\ref{fig:7}. Since, the larger the
depletion is, the smoother the electron density is. Remarkably, the depletion length has a
major influence on the amplitude and location  of the resistance \emph{dip}.
To summarize, we have shown that the \emph{dip} effect observed in the Hall
resistance under the quantized Hall effect
conditions is strongly influenced by the sample parameters and can be well explained
by the self-consistent screening theory. We found that, the \emph{dip} relates strongly on sample width and can be commonly observed in large samples. Our calculations show that if the edge effects are predominant the resistance \emph{dip} effect increases. In addition, for the large samples disorder effects become more important. As a numerical check we also performed calculations using different number of mesh points and observed that for sufficiently high number of mesh points (501) the results are quantitatively unaffected.

\section{Experimental realization: Proof of concept}
In our final investigation we performed standard Hall resistance measurements on narrow ($W$=10 $\mu$m) Hall bars, which are defined by metallic gates on the surface. The nominal electron mobility is $8.0\times10^6$ Vs/cm$^{-2}$ and 2DEG resides some 100 nm below the surface. A negative gate voltage $V_g$ is applied to deplete the electron gas below the surface, where the pinch-off voltage is estimated to be $V_g=-0.3$ V. Experiments are conducted in a dilution fridge equipped with a 20 T super-conducting magnet at 100 mK. An AC voltage is driven between the source and drain contacts at 11.5 Hz, with an excitation amplitude of 100 mV corresponding an sample current of 100 nA. Details of experimental setup and sample properties are discussed elsewhere~\cite{metin:2013}.
\begin{figure}
\resizebox{0.5\textwidth}{!}{%
  \includegraphics{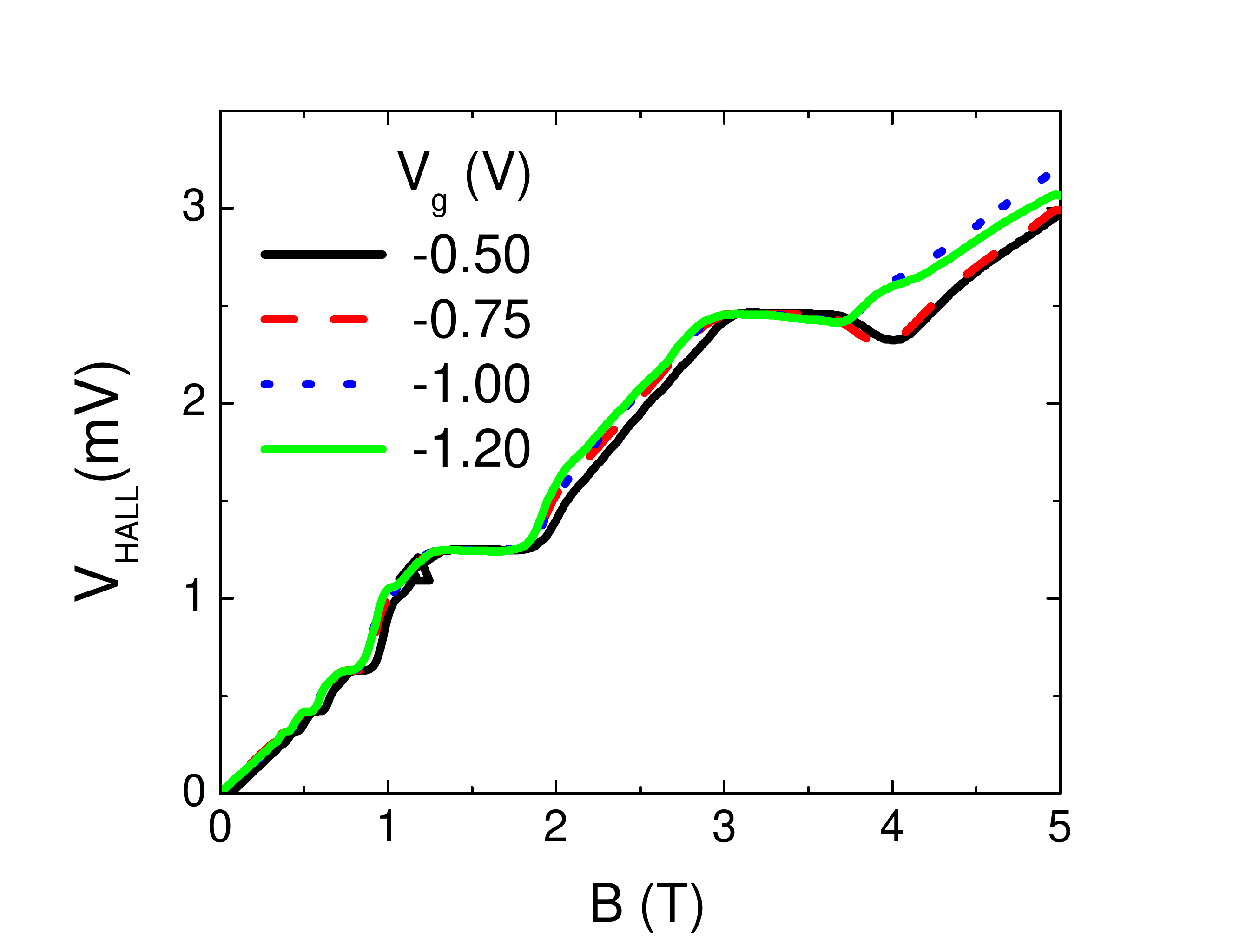}
}
\caption{Hall resistance measured as a function of magnetic field considering different gate voltages.}
\label{fig:8}       
\end{figure}
Fig.~\ref{fig:8} depicts a typical Hall resistance measurement, where the gate potential is changed to manipulate the edge profile. The first observation is that the average electron density remains unaffected by the change of $V_g$, as can be seen from the slope of the Hall voltage at low $B$ field. The dip effect is observed at the high-field end of the $\nu=2$ plateau when low gate voltages are applied, however, which are sufficiently larger than the pinch-off voltage. Once the gate voltage is increased, the dip effect tends to disappear as predicted by our numerical calculations. A systematic experimental investigation is an ongoing project, where the temperature, sample width and mobility effects are also studied. However, our preliminary findings as shown here, strongly supports our model.
\section{Conclusion}
\label{conc}

In this work we investigated the unexpected Hall resistance decrease, namely the dip effect, utilizing the screening theory of the quantized Hall effect. We observed that, at sufficiently narrow samples the effect under consideration tends to disappear while the edge electrostatics allow the formation of wide incompressible channels. Once the width of the incompressible strip becomes comparable with the magnetic length, the strip becomes unstable and collapses. The constraining conditions of the collapse is determined by the steepness of the confinement potential, the strength of direct Coulomb interaction and the Lorentz force acting on the electrons at the edges. Numerically investigating the strip widths we showed that at  wide samples incompressible strips can collapse and be reconstructed yielding a dip effect. The effect itself is immune to level broadening, however, the long-range fluctuations may shift the center of the dip, depending on the parity of the modulation period. We also showed that temperature effects has an important influence on the effect, while the formation and the stability of the incompressible strips are essentially determined by temperature. In the next step, we also considered the effect of depletion length which essentially describes the strength of the confinement potential at the edges. We observed that, wider depleted regions yield both a shift in the dip and also increase the amplitude of the effect. Experimental investigations in light of our calculations provided a reasonably well agreement between theory and measurements. A deeper understanding of the dip effect can be achieved by performing systematic experiments, which is an on going project.

\section{Acknowledgments}

This work was partially supported by the Istanbul University under grant  IU-BAP:6970 and IU-BAP: 22662. S.B., S.S. and E.M.K. acknowledges financial support from the T\"UBiTak under grant 112T264 and 211T148.

%

\begin{thebibliography}{}
%
%














\bibitem{Klitzing}
K. v. Klitzing, G. Dorda and M. Pepper, Phys. Rev. Lett. \textbf{45}, (1980) 494.

\bibitem{Siddiki04:195335}
A.~Siddiki and R.~R. Gerhardts, Phys. Rev. B \textbf{70}, (2004) 195335.

\bibitem{Chklovskii92:4026}
D.~B. Chklovskii, B.~I. Shklovskii, and L.~I. Glazman, Phys. Rev. B \textbf{46}, (1992) 4026.

\bibitem{Lier}
K.~Lier and R.~R. Gerhardts, Phys. Rev. B \textbf{50}, (1994) 7757.

\bibitem{Siddiki:2003}
A. Siddiki, R. R. Gerhardts,  Phys. Rev. B \textbf{68}, (2003) 125315:1.

\bibitem{Guven}
K. Guven and R. R. Gerhardts, Phys. Rev. B \textbf{ 67}, (2003) 115327.

\bibitem{Ando}
T. Ando, A. B. Fowler and F. Stern, Rev. Mod. Phys. \textbf{54}, (1982) 437.

\bibitem{Gerhardts}
R. R. Gerhardts, Phys. stat. sol. (b) \textbf{245}, (2008) 378.

\bibitem{Shlimak:2006}
I.~Shlimak, K.-J.~Friedland, V.~Ginodman and S.V.~Kravchenko,  phys. stat. sol. (c) \textbf{3}, (2006) 309.

\bibitem{Galistu}
G.~Galistu, {\it Academisch Proefschrift} (2010) 1-127.

\bibitem{Shlimak:prb2006}
I.~Shlimak,  V.~Ginodman, K.-J.~Friedland and S.V.~Kravchenko, Phys. Rev. B \textbf{73}, (2006) 205324.

\bibitem{Richter}
C. A. Richter, R. G. Wheeler and R. N. Sacks, Surf. Sci. \textbf{263}, (1992) 270.

 \bibitem{Griffin}
N. Griffin, R. B. Dunford, M. Pepper, D. J. Robbins, A. C. Churchill and W. Y. Leong, J. Phys: Condens. Mater.
\textbf{12,} (2000) 1811.

\bibitem{Ramvall}
P. Ramvall, N. Carlsson, P. Omling, L. Samuelson, W. Seifert, Q. Wang,
K. Ishibashi and Y. Aoyagi, J. Appl. Phys. \textbf{84}, (1998) 2112.

\bibitem{Shlimak:2005}
I. Shlimak, V. Ginodman, A. B. Gerber, A. Milner, K.J. Friedland and D. J. Paul, Europhys. Lett.
\textbf{69}, (2005) 997.

\bibitem{Siddiki:2010}
A. Siddiki, S. Erden Gulebaglan, N. Boz Yurdasan, G. Bilgec, A. Yildiz and I. Sokmen, Europhys. Lett. \textbf{92}, (2010) 67010.

\bibitem{metin:2013}
E. M. Kendirlik, S. Sirt, S. B. Kalkan, W. Dietsche, W. Wegscheider, S. Ludwig and A. Siddiki, arXiv:1305.1156, (2013).

\bibitem{sailer:2010}
J. Sailer, A. Wild, L. Lang, A. Siddiki and D. Bougeard, New Journal of Physics \textbf{12}, (2010) 113033.

\bibitem{jose:2008}
J. Horas, A. Siddiki, J. Moser, W. Wegscheider, and S. Ludwig, Physica E:Low-dimensional Systems and Nanostructures \textbf{40}, (2009) 1130.

\bibitem{siddiki:2009}
A. Siddiki, J. Horas, J. Moser, W. Wegscheider, and S. Ludwig, Europhys. Lett. \textbf{88}, (2009) 17007.

\bibitem{siddiki:2010njp}
A. Siddiki, J. Horas, D. Kupidura, W. Wegscheider, and S. Ludwig, New Journal of Physics \textbf{12}, (2010) 113011.

\bibitem{Sinem:2012}
S. Erden Gulebaglan, G. Oylumluoglu, U. Erkaslan, A. Siddiki and I. Sokmen, Physica E:Low-dimensional Systems and Nanostructures \textbf{44}, (2012) 1495.
































\end{thebibliography}
%

\end{document}